\documentclass[12pt]{amsart}
\usepackage[active]{srcltx}
\usepackage[all
]{xy} \xyoption{poly}

\usepackage{xspace}
\usepackage{enumitem}
\usepackage[mathscr]{eucal}
\usepackage{amssymb}
\usepackage{a4wide}
\usepackage{ifthen}
\usepackage{amscd}
\usepackage[hyperindex,bookmarksnumbered,
plainpages]{hyperref}

\def\X{\bf X}
\def\x{\bf x}
\def\Y{\bf Y}
\def\y{\bf y}
\def\Z{\bf Z}
\def\z{\bf z}
\def\u{\bf u}
\def\U{\bf U}

\newtheorem{thrm}{Theorem}
\newtheorem{thm}{Theorem}[section]
\newtheorem{lem}[thm]{Lemma}
\newtheorem{rem}[thm]{Remark}

\def\O{\mathbb O}
\def\R{\mathbb R}

\def\C{\mathbb{C}}

\def\R{\mathbb{R}}

\def\Proof{{\it Proof.\ }\ }

\newcommand{\be}{\begin{equation}}
\newcommand{\ee}{\end{equation}}
\newcommand{\bes}{\begin{eqnarray*}}
\newcommand{\ees}{\end{eqnarray*}}
\newcommand{\bee}{\begin{eqnarray}}
\newcommand{\eee}{\end{eqnarray}}

\def\ctd{\hfill$\Box$}

\theoremstyle{remark}

\makeindex

\begin{document}

\title{ Triple Jordan systems  and \\[3mm] integrable  models of mKdV-type}

\medskip

\author{I.\,P.\,Shestakov and V.\,V.\,Sokolov} 

\maketitle
\begin{abstract} A one-to-one correspondence between triple Jordan systems and integrable multi-component models of the modified Korteveg--de Vries type is established. 
\end{abstract}
\pagestyle{plain}

\section{\Large \bf Introduction.} 

 One of the most remarkable observations by Sergey Svinolupov is the
discovery of the fact that polynomial multi-component integrable equations 
are closely related to the well-known non-associative algebraic structures  such 
as left-symmetric algebras, Jordan algebras,  triple Jordan systems, etc.
This connection allows to clarify the nature of known vector and
matrix generalizations (see, for instance, \cite{For1,For2,For3}) of classical
scalar integrable equations  and to construct some new examples of this
kind \cite{svsok}.

One of his results is related to integrable multi-component generalizations of the celebrated Korteweg-de Vries equation 
$u_t = u_{xxx}+ 6 u u_x$. This equation was integrated by means of the inverse scattering method \cite{AblSeg81}. Several algebraic structures are associated with this equation. In particular, the KdV equation has infinitely many infinitesimal symmetries (or  commuting flows). The existence of higher symmetries  is a foundation stone of the symmetry approach to classification of integrable systems (see, for example,  \cite{ss, mss, miksok}).

In the paper \cite{svin3} systems of form 
\begin{equation} \label{kort}
u^i_t=u^i_{xxx}+\sum_{k,j}\, C^i_{jk} \,u^k \, u^j_x, \qquad i,j,k=1,\dots,N
\end{equation}
have been considered. Let us associate with the system \eqref{kort} the $N$-dimensional algebra ${\mathcal A}$ with the structural constants $ C^i_{jk}$. Denote the product in ${\mathcal A}$ by $\circ.$

\begin{thrm}\label{thm3}  {\rm \cite{svin3}}. Suppose that the algebra ${\mathcal A}$ is commutative. Then system 
\eqref{kort} has a polynomial symmetry of order  $m \ge 5$  iff ${\mathcal A}$ is a { \it Jordan algebra}.
\end{thrm}

The original proof of Theorem \ref{thm3} was obtained by straightforward computations in terms of structural constants. It turns out 
that the computations can be performed in terms of algebraic operations, which define the equation and the symmetry. We demonstrate the corresponding technique in Section 2.

If we do not assume that the algebra ${\mathcal A}$ is commutative, then a generalization of Theorem \ref{thm3} is given by the following  

\begin{thrm}\label{thm4} System 
\eqref{kort} has a polynomial symmetry of order  $m \ge 5$ iff 
\begin{itemize}
\item The identity $$
(X\circ Y-Y\circ X)\circ Z = 0
$$
holds in $\mathcal A.$  In this case the vector space $J = [\mathcal A,\, \mathcal A]$ is a double-side ideal in $\mathcal A.$
\item The quotient algebra ${\mathcal A}/J$ is a Jordan one.
\end{itemize}
\end{thrm}
We did not find this result in Svinolupov's papers. 

In \cite{svin2} systems of form
\begin{equation}\label{jormkdv}
u^i_t=u^i_{xxx}+\sum_{j,k,m} B^i_{jkm} u^j u^k u^m_x, \qquad i,j,k=1,\dots,N
\end{equation}
were considered. They are multi-component generalizations of the modified KdV equation 
\begin{equation}\label{scmkdv}
u_t=u_{xxx}+6 u^2 u_x, \qquad   u=u(x, t).
\end{equation}
Equation \eqref{scmkdv} is known to be integrable. In particular, it has 
infinitely many infinitesimal symmetries. 

There exists the following integrable matrix generalization 
\begin{equation}\label{matmkdv1}
{\bf U}_{t}={\bf U}_3+3 {\bf U}^2 {\bf U}_1+3 {\bf U}_1 {\bf U}^2
\end{equation}
of equation \eqref{scmkdv}.
 Here $U(x,t)$ is a matrix of arbitrary size $m \times m.$ Written in terms of components of the matrix $U$, this system belongs to the class of systems of  form \eqref{jormkdv}.
For system \eqref{matmkdv1} we have $N=m^2.$

The main observation made by Svinolupov is that for any triple Jordan system with the structural constants $B^i_{jkm}$ the corresponding system \eqref{jormkdv} has infinitesimal symmetries. This statement was not proved in \cite{svin2}. For other relations between integrable models and triple Jordan systems see \cite{svin1}.

In the next section we prove Svinolupov's statement. As our main result, we also formulate and prove a converse assertion: if a system \eqref{jormkdv} has symmetries, then 
$$B^i_{jkm}=S^i_{jmk}+S^i_{mjk},$$ 
where $S^i_{jkm}$ are structural constants of a triple Jordan system. 

The class of systems \eqref{jormkdv} is invariant with respect to the group of linear transformations of the variables $u^1,\dots , u^N$. A system \eqref{jormkdv} is called {\it reducible} if it can be reduced to a system of the form
\begin{equation}\label{red}
\left\{
\begin{array}{ll}
u^i_t=  u^i_{xxx}+\sum_{j,k,m} B^i_{jkm} u^j u^k u^m_x, \ j,k,m, i=1,\dots, l, \\[2mm]
u^i_t=   u^i_{xxx}+\sum_{j,k,m} B^i_{jkm} u^j u^k u^m_x, \ j,k,m=1,\dots,N,\ i=l+1,\dots, N.
\end{array}
\right.
\end{equation}

In other words, a system is reducible if it has a subsystem of the same form but of lower dimension. In Section 2 we prove (cf. \cite{svin2}) that a system \eqref{jormkdv} is irreducible iff the corresponding Jordan triple system is simple. A similar statement for systems of form \eqref{kort} was proved in \cite{svin3}.
  
\section{\Large \bf MKdV type systems.}

The modified Korteweg-de Vries equation \eqref{scmkdv}
is one of most celebrated equations solved by the inverse scattering method. This equation possesses infinitely many higher (infinitesimal) symmetries of odd orders.

A higher (or generalized) infinitesimal symmetry  of evolution equation of the form  
\begin{equation}\label{eveq}
u_t=F(u, u_x, u_{xx}, ... , u_{n}), \qquad  u_i=\frac{\partial^i u}{\partial x^i},
\end{equation}
is an evolution equation
\begin{equation}\label{evsym}
u_{\tau}=G(u, u_x,  u_{xx}, \dots , u_m), \qquad m > 1, 
\end{equation}
which is compatible with~\eqref{eveq}. Compatibility means that \begin{equation}\label{ttau}
\frac{\partial}{\partial t}\frac{\partial u}{\partial \tau}=\frac{\partial}{\partial \tau}\frac{\partial u}{\partial t},
\end{equation}
where the partial derivatives are calculated in virtue of~\eqref{eveq} and~\eqref{evsym}. For rigorous definition consider the ring 
${\mathcal F}$ of polynomials that depend of finite number of independent variables $u,u_1, u_{2}, \dots$. 
As usual in differential algebra, we have a principle derivation \begin{equation} \label{DD}
D \stackrel{def}{=}  \sum_{i=0}^\infty u_{i+1} \frac{\partial}{\partial u_i},
\end{equation}
 which generates all independent variables $u_i$ starting from $u_0=u$. We associate with equation \eqref{eveq} the infinite-dimensional vector field   
\begin{equation}
\label{Dt} D_F= \sum_{i=0}^\infty D^{i}(F) \frac{\partial}{\partial u_i},
\end{equation}
This vector field commutes with $D$. We call vector fields of form \eqref{Dt} {\sl evolutionary}. The set of all evolutionary vector fields is a Lie algebra over $\C.$
  By definition, the compatibility of~\eqref{eveq} and~\eqref{evsym} means that the vector fields $D_F$ and $D_G$ commute.

Equation (\ref{eveq}), where $F$ is a polynomial, 
 is said to be $\lambda $-{\it homogeneous} of {\it order} $\mu $ if it 
admits the one-parameter group of scaling symmetries
$$(x, \ t, \ u) \mapsto  (\varepsilon^{-1}x, \ \varepsilon^{-\mu} t, \ \varepsilon^{\lambda} u).$$
For $N$-component systems with unknowns $u^1,...,u^N$ the corresponding 
scaling group has a similar form
\begin{equation}\label{homo}(x,t,u^1,...,u^N)\mapsto (\varepsilon^{-1} x, \ \varepsilon^{-\mu} t, \ 
\varepsilon^{\lambda_1} u^1,...,
\varepsilon^{\lambda_N} u^N).
\end{equation}
Equation (\ref{scmkdv}) is homogeneous with $\mu=3, \lambda=1$ and its simplest symmetry 
\begin{equation}\label{msym}
u_{\tau}=u_5+10 u^2 u_3+40 u u_1 u_2+10 u_1^3+30 u^4 u_1.
\end{equation}
is homogeneous with  $\mu=5, \lambda=1.$ Equations  (\ref{scmkdv}) and  (\ref{msym})  are also invariant with respect to the discrete involution $u\to -u.$  It was proved in \cite{sw} that if a $\lambda$-homogeneous third order scalar equation with $\lambda \ne 0$ has infinitely many symmetries, then it has a symmetry of fifth order.

Let $B$ be a  triple system with basis ${\bf e}_1,...,{\bf e}_N,$ such that 
$$
\{{\bf e}_j,{\bf e}_m,{\bf e}_k\} =\sum_i B^i_{jkm} {\bf e}_i.
$$
If $\, U=\sum_k u^k {\bf e}_k, \,$ then 
the algebraic form of the system (\ref{jormkdv}) is given by
\begin{equation} \label{mkdvsvin}
U_t = U_{xxx} + 3 B(U, U_x, U)
\footnote{We put the coefficient 3 here and 5 below in \eqref{symmkdvsvin} to avoid rational numbers in formulas for the symmetries.}.
\end{equation}
Triple systems $B(X,Y,Z)$ such that  $B(X,Y,Z) = B(Z,Y,X)$  are in one-to-one correspondence with systems of type \eqref{jormkdv}. Actually, a triple system $B$ is defined up to a constant factor, which corresponds to the scaling $U\to {\rm const}\, U.$

Recall that a subspace $I$ of a triple system $B$ is called {\em an ideal} if $\{I,B,B\}+\{B,I,B\}+\{B,B,I\}\subseteq I$.

\begin{lem}  
A system of form (\ref{jormkdv}) is reducible if and only if  the corresponding triple system $B(X,Y,Z)$ has a non-trivial ideal $I$. In this case an independent subsystem  corresponds to the quotient-system $B/I$.
\end{lem}
{\Proof}
Assume first that a system $B$ may be reduced to form (\ref{red}). Let $I$ be the subspace spanned by $e_{l+1},\ldots,e_N$. If at least one of the indices $j,k,m$ is more than $l$ then $B_{jkm}^i=0$ for all $i<l$ and $\{e_j,e_k,e_m\}\in I$, which proves that $I$ is an ideal of $B$.

\smallskip
Conversely, assume that $B$ has an ideal $I$. Choose in $B$ a basis $e_1,\ldots, e_l,e_{l+1},\ldots,e_N$ such that the last $N-l$ elements form a basis of $I$. Let us write $ U=\sum_{i=1}^N u^i {\bf e}_i$ in the form $U=V+W$, where $V=\sum_{i=1}^l u^i {\bf e}_i$ and $W=\sum_{i=l+1}^N u^i {\bf e}_i$. Then $W\in I$ and we have
\bes
U_t&=&U_{xxx}+B(U,U_x,U)=V_{xxx}+B(V,V_x,V)\\
&+&W_{xxx}+B(W,V_x+W_x,V+W)+B(V,W_x,V+W)+B(V,V_x,W).
\ees
We have $B(V,V_x,V)=B(V,V_x,V)|_V+B(V,V_x,V)|_W$,  where $B(V,V_x,V)|_V\in V$, $B(V,V_x,V)|_W\in W$. Now our system is reduced to the following subsystems of form (\ref{red}):
\bes
V_t&=&V_{xxx}+ B(V,V_x,V)|_V,\\
W_t&=&W_{xxx}+ B(V,V_x,V)|_W\\
&+&B(W,V_x+W_x,V+W)+B(V,W_x,V+W)+B(V,V_x,W).
\ees
It is also clear that the independent subsystem $V$ corresponds to the quotient triple system $B/I$.

\ctd

Therefore, irreducible systems correspond to simple  triple systems $B(X,Y,Z)$.

\smallskip

Equations \eqref{mkdvsvin} are homogeneous under transformations \eqref{homo} with $\mu=3, \lambda=1$ 
and invariant with respect to the discrete involution $U\mapsto -U.$ Without loss of generality we assume that all polynomial symmetries enjoy the same properties.
Indeed, if \eqref{mkdvsvin} has a polynomial symmetry then any homogeneous component of its right hand side define a symmetry and we may consider only homogeneous symmetries. Similarly, both parts of a polynomial symmetry, symmetric and skew-symmetric under the involution $U\to -U,$ are symmetries. That is why we assume that the symmetry does not contain terms of even degrees.

By analogy with the scalar case we are looking for a fifth order symmetry for \eqref{mkdvsvin}.  Under conditions described above such  symmetry is given by 
\begin{equation}\label{symmkdvsvin}
U_{\tau}=U_{5}+5 B_1(U,U,U_3)+5 B_2(U,U_1,U_2)+5 B_3(U_1,U_1,U_1)+5 C(U,U,U,U,U_1),
\end{equation}
where $B_i$ are some triple systems and $C$ is a 5-system.  

\smallskip

In \cite{svin2} the following statement was formulated: 

\begin{thrm}\label{thm1}  For any  triple Jordan system $\{\cdot,\cdot,\cdot\}$  equation \eqref{mkdvsvin}, where $B(X,Y,X)=\{X,X,Y \},$  has a fifth order symmetry of form 
\eqref{symmkdvsvin}. 
\end{thrm}

The original (unpublished) proof of Theorem \ref{thm1} was obtained by straightforward computations in terms of structural constants of the operations 
$B,B_1,B_2,B_3$   and $C$. It turns out 
that the computations can be performed in terms of these algebraic operations  and identities, which relate them. This drastically simplifies the proof. 
 
\smallskip

In this section we prove a bit stronger statement. 

\begin{thrm}\label{thm2} Equation \eqref{mkdvsvin} has a fifth 
order symmetry of form \eqref{symmkdvsvin}
iff
\begin{equation}\label{BBB}
B(X,Y,Z)=\{X,Z,Y\}+\{Z,X,Y\},
\end{equation}
where $\{\cdot,\cdot,\cdot\}$ is a triple Jordan system.
\end{thrm}
{\Proof} The compatibility condition
\begin{equation}\label{com}
0=(U_t)_{\tau}-(U_{\tau})_t=P(U,U_1,...,U_5)
\end{equation}
of (\ref{mkdvsvin}) and (\ref{symmkdvsvin}) 
leads to a differential polynomial $P$ that should be identically zero. After the scaling $U_i \mapsto z_i U_i$ in $F$ all coefficients of different monomials in $z_0,..., z_5$ have to be identically equal to zero. Equating the coefficient of $z_0 z_1 z_5$ to zero, we find that 
\begin{equation}\label{B1}
B_1(X, X, Y)=B(X, Y, X).
\end{equation}
The coefficient of $z_0 z_2 z_4$ leads to 
\begin{equation}\label{B2}
B_2(X, Y, Z)= 2 B(X, Y, Z)+2 B(X, Z, Y).
\end{equation}
All other terms containing  $z_5$ and $z_4$ disappear by virtue of \eqref{B1} and \eqref{B2}.
Comparing the coefficients of $z_1 z_2 z_3$, we obtain 
\begin{equation}\label{B3}
B_3(X, X, X)= B(X, X, X),
\end{equation}
while the coefficients of $z_3 z_1 z_0^3$ give rise to 
\begin{equation}\label{CC}
C(X, X, X, X, Y)= B(X, B(X,Y,X), X) + \frac{1}{2} B(X,Y,B(X,X,X)).
\end{equation}
Thus the symmetry \eqref{symmkdvsvin} is expressed in terms of the triple system $B.$ All fifth order identities $I_i=0, \,i=1,2,3,4$ for the triple system $B$ come from the coefficients of $z_0^3 z_1 z_3, z_0^3 z_2^2, z_0^2 z_1^2 z_2$ and $z_0 z_1^4$. Here
\bes
I_1(X,Y,Z)&=&2 B(X,Z,B(X,X,Y))-3 B(X,Z,B(X,Y,X))+B(Y,Z,B(X,X,X)),\\[2mm]
I_2(X,Y)&=&2 B(X,Y,B(X,X,Y))-3 B(X,Y,B(X,Y,X))+B(Y,Y,B(X,X,X)),\\[2mm]
I_3(X,Y,Z)&=&2 B(X, Y, B(X, Y, Z)) - 6 B(X, Y, B(X, Z, Y)) +2 B(X, Y, B(Y, X, Z))   \\
&-&4 B(X, Z, B(X, Y, Y))+2 B(X, Z, B(Y, X, Y)) - 
 2 B(X,  B(Y, Z, Y), X)\\
& +& 2 B(Y, Y,  B(X, X, Z))  -3 B(Y, Y,  B(X, Z, X)) + 4 B(Y, Z, B(X, X, Y))\\
 &-& 2 B(Y, Z, B(X, Y, X)) + 4 B(Y, B(X, Y, X), Z) +  2 B(Y, B(X, Z, X),Y)\\
 &+& 2 B(Z, Y, B(X, X, Y)) -  3 B(Z, Y, B(X, Y, X)),
 \ees
and
\bes
I_4(X,Y)&=&B(X, Y, B(Y, Y, Y)) + 2 B(Y, Y, B(X, Y, Y))\\
 &-& B(Y, Y, B(Y, X, Y)) - 
 2 B(Y, B(X, Y, Y), Y).
\ees
It is clear that $I_2(X,Y)=I_1(X,Y,Y).$ Using  the method of undetermined coefficients, we will show that the identity $I_4=0$ is a consequence of the identities $I_2=0$ and $I_3=0.$ First, introduce the polarizations of these identities. 
Let \begin{itemize} \item $J_2(X,Y,Z,U,V)$ be the coefficient of $k_1 k_2 k_3 $ in $I_2(k_1 X + k_2 U + k_3 V, Y, Z)$; \\
\item $J_3(X,Y,Z,U,V)$ be the coefficient of $k_1 k_2 k_3 k_4$ in $I_3(k_1 X + k_2 U, k_3 Y + k_4 V, Z)$; \\
\item $J_4(X,Y,Z,U,V)$ be the coefficient of $k_1 k_2 k_3 k_4$ in $I_4(X, k_1 Y + k_2 Z + k_3 U + k_4 V)$.
\end{itemize}
 Consider the following expression
 $$
 \begin{array}{c}
 Z=J_4(X,Y,Z,U,V)-\sum_{\sigma\in S_5} b_{\sigma} J_2\Big(\sigma(X),\sigma(Y),\sigma(Z),\sigma(U),\sigma(V)\Big)-\\[2mm]
  \sum_{\sigma\in S_5} c_{\sigma} J_3\Big(\sigma(X),\sigma(Y),\sigma(Z),\sigma(U),\sigma(V)\Big),
 \end{array}
 $$
 where $\sigma$ is a permutation of the set $\{X,Y,Z,U,V\}.$ To take into account the identity $B(X,Y,Z)=B(Z,Y,X),$ we fix the ordering
 $$
 U<V<X<Y<Z< B(\cdot,\cdot,\cdot)
 $$
 and replace all expressions of the form $B(P,Q,R)$ by  $B(R,Q,P)$ if $P>R.$ After that, equating the coefficients of similar terms in the relation $Z=0$, we obtain an overdetermined system of linear equations for the coefficients $b_{\sigma}$ and $c_{\sigma}$. Solving this system, we find that
\bes
&& J_4(X,Y,Z,U,V)=\\
&=&\frac{1}{6}\Big(J_2(U, X, V, Y, Z) + J_2(U, X, Y, V, Z) + J_2(U, X, Z, V, Y) + 
 J_2(V, X, U, Y, Z) \\
&-& J_3(U, V, X, Y, Z)-J_3(U, V, X, Z, Y)-J_3(U, V, Y, X, Z)- J_3(U, V, Z, X, Y)\\
&-& J_3(U, Y, V, X, Z)-J_3(U, Y, X, V, Z)-J_3(V, U, X, Y, Z)-J_3(V, U, X, Z, Y)\\
&-& J_3(V, U, Y, X, Z) - J_3(V, U, Z, X, Y)- J_3(V, Y, U, X, Z) - J_3(X, U, V, Y, Z)\\
 &-& J_3(X, U, V, Z, Y)- J_3(X, U, Y, Z, V) - J_3(X, U, Z, Y, V)  - J_3(X, V, U, Y, Z)\\
 &-&J_3(X, V, U, Z, Y) - J_3(Y, U, X, Z, V)\Big).
 \ees

Consider a triple system 
\bee\label{JTS}
\{X,Y,Z\} = \frac{1}{2} \Big( B(Y,Z,X)+B(Y,X,Z)-B(X,Y,Z)  \Big).
\eee
It is easy to verify that
\begin{equation}\label{BB}
B(X,Y,Z)= \{X,Z,Y\} + \{Z,X,Y\}.
\end{equation}
Let us prove that the identities $J_2=J_3=0$ are equivalent to the fact that the triple system $ \{X,Y,Z\}$ is Jordan, that is, satisfies the identity
\bes
\{X, Y, \{U, V, Z\}\} - \{\{X, Y, U\}, V, Z\} - 
 \{U, V, \{X, Y, Z\}\} + \{U,\{ Y, X, V\}, Z\}=0.
\ees
Let us rewrite the left side of this identity in terms of the triple system $B$ by means of \eqref{JTS} and denote the result by  ${\mathcal J}$.
 By the same method of undetermined coefficients we verified that the identity ${\mathcal J}=0$ follows from $J_2=J_3=0$ and vice versa,
 each of the identities $J_2$ and $J_3=0$ follows from ${\mathcal J}=0.$ For example, 
 $$
 \begin{array}{c}
J_2(X,Y,Z,U,V) = -{\mathcal J}(U, X, V, Z, Y) + {\mathcal J}(U, X, Y, Z, V) - {\mathcal J}(U, Y, X, Z, V) + \\[1.5mm]
 {\mathcal J}(V, X, Y, Z, U) - {\mathcal J}(V, Y, U, Z, X) - {\mathcal J}(V, Y, X, Z, U) - \\[1.5mm]
 {\mathcal J}(X, U, V, Z, Y) + {\mathcal J}(X, U, Y, Z, V) - {\mathcal J}(X, V, U, Z, Y) + \\[1.5mm]
 {\mathcal J}(Y, U, V, Z, X) + {\mathcal J}(Y, V, U, Z, X) + {\mathcal J}(Y, V, X, Z, U).
 \end{array}
 $$
 The formulas, which express $J_3$ through ${\mathcal J}$ and ${\mathcal J}$ through $J_2, J_3$, are more complicated.
 
 Besides the above fifth order identities there exist two identities of order 7. The coefficient of $z_0^6 z_2$ in the polynomial $P$ yields 
 $$
 B(X, B(X, Y, X),\, B(X, X, X)) - B(X, B(X, Y, B(X, X, X)), X)=0
 $$
 while the coefficient of $z_0^5 z_1^2$ leads to
 
 \bes
 &&2 B[X, Y, B[X, X, B[X, Y, X]]] - 2 B[X, Y, B[X, Y, B[X, X, X]]] \\
 &-&  3 B[X, Y, B[X, B[X, Y, X], X]] +
  4 B[X, B[X, Y, X], B[X, Y, X]] \\
  &+& 
 2 B[X, B[X, Y, Y], B[X, X, X]] - 2 B[X, B[X, Y, B[X, X, Y]], X] \\
 &+&3 B[X, B[X, Y, B[X, Y, X]], X] - 4 B[X, B[X, B[X, Y, X], Y], X]  \\
 &-&B[X, B[Y, Y, B[X, X, X]], X] + B[B[X, X, X], Y, B[X, Y, X]]=0.
 \ees
 
 Using the method of undetermined coefficients, one can check that both identities follow from ${\mathcal J}=0$.  Thus we verified that  all identities, which are produced by the compatibility condition \eqref{com} are equivalent to \eqref{B1}-\eqref{CC} and ${\mathcal J}=0$.
\ctd
\smallskip
 
\begin{rem} Since equation \eqref{mkdvsvin} is expressed via $B(X,Y,X)$, it follows from \eqref{BB} that all equations that have the fifth order symmetry are described by Theorem \ref{thm1}.
\end{rem}
\smallskip

Formulas (\ref{JTS}), (\ref{BB}) show that the triple system $B(X,Y,Z)$ is simple if and only if the corresponding Jordan triple system $\{X,Y,Z\}$ is simple. 
According to the classification of simple Jordan triple systems (see, for example, \cite{Zel}), we may now give the following examples of irreducible integrable  vector mKdV systems admitting fifth order symmetries:

\begin{enumerate}
\item The triple Jordan system defined on the set of all $m\times m$ matrices by  the operation
\bes
\{\X,\Y,\Z\}=\X\Y\Z+\Z\Y\X
\ees
gives the matrix mKdV equation \eqref{matmkdv1}.

\item
Let $V=\R^n$ be a Euclidean vector space with the scalar product $(x,y)$. The triple Jordan product on $V$  defined by 
\bes
\{\x,\y,\z\}=(\x,\y)\z+(\z,\y)\x-(\x,\z)\y
\ees
gives a the mKdV system
\bes
\u_t=\u_{xxx}+3(\u,\u)\u_x.
\ees
\item
The space of rectangular $m\times n$ matrices $M_{m,n}(\R)$ with  the Jordan product $\{\X,\Y,\Z\}=\X\Y^t\Z+\Z\Y^t\X$ defines a  mKdV system
\bes
\U_t=\U_{xxx}+3(\U\U^t\U_x+U_x\U^t\U).
\ees
\item 
The spaces of symmetric (skew-symmetric) $m\times m$ matrices are closed with respect to the triple Jordan matrix product in (1) and define mKdV systems of dimensions $m(m+1)/2,\, m(m-1)/2$, respectively.

Similarly, the subspaces of Hermitian and skew-Hermitian $2m\times 2m$ matrices with respect to the symplectic involution define mKdV systems of dimensions $m(2m-1)$ and $2m(m-1)$ respectively. 
\item
The space of $1\times 2$ matrices $M_{1,2}(\O)$ over the algebra of octonions $\O$ with the triple Jordan product $\{\X,\Y,\Z\}=(\X\Y^t)\Z+(\Z\Y^t)\X$ defines a 16-dimensional mKdV system.
\item
The 27-dimensional exceptional simple Jordan algebra $H(\O_3)$ of $3\times 3$ hermitian matrices over octonions ({\em the Albert algebra}) defines a 27-dimensional mKdV system with respect to the triple Jordan product 
$$
\{\X,\Y,\Z\}=(\X\cdot \Y)\cdot \Z+(\Z\cdot \Y)\cdot \X-(\X\cdot \Z)\cdot \Y,
$$
 where $\X\cdot \Y=\frac12 (\X\Y+\Y\X)$.
\end{enumerate}

\section{Acknowledgments} 

The authors are grateful to A. Sevostyanov for useful discussions. 
The research was carried out during a stay of the second author at the University of S\~ao Paulo, supported by the FAPESP grant 2016/07265-8.
The author  thanks FAPESP for the support and the Institute of Mathematics of the University of S\~ao Paulo for providing excellent working conditions.
The first author acknowledges the supports by FAPESP grant 2014/09310-5 and CNPq grant 303916/2014-1.

\end{document}